\documentclass[12pt]{article}
\pdfoutput=1

\global\arraycolsep=1pt
\usepackage{amssymb,amsmath,graphicx,color,mathrsfs}
\usepackage {pstricks}
\usepackage[cmbtt]{bold-extra}
\usepackage[font={small,it},labelfont=bf]{caption} 
\usepackage{listings}

 \usepackage{float}

\usepackage[
      colorlinks=true,
      linkcolor=myblue,
      urlcolor=myblue,
      filecolor=myblue,
      citecolor=red,
      pdfstartview=FitV,
      ]{hyperref}
\usepackage{cite}

\renewcommand{\thefootnote}{\fnsymbol{footnote}}

\usepackage{wrapfig}

\definecolor{myblue}{rgb}{0.2,0.2,0.7}
\definecolor{myred}{rgb}{0.9,0.2,0.1}
\definecolor{dayflower}{cmyk}{0.9,0.36,0,0}%blue
\definecolor{LightBlue}{cmyk}{0.45,0.04,0.14,0}
\definecolor{YellowishGreen}{cmyk}{0.56,0.08,0.95,0}%green
\definecolor{siskin}{cmyk}{0.28,0.1,0.95,0}%yellow
\definecolor{YoungLeaves}{cmyk}{0.35,0,0.47,0.04}
\definecolor{bark}{cmyk}{0.15,0.2,0.92,0}
\definecolor{kariyasuiro}{cmyk}{0.18,0.1,0.95,0}
\definecolor{BushWarbler}{cmyk}{0.37,0.42,0.82,0.65}
\definecolor{FadedVermilion}{cmyk}{0,0.7,0.73,0.1}
\definecolor{Peony}{cmyk}{0.18,0.82,0,0}%red
\definecolor{ThinChineseInk}{cmyk}{0.08,0.05,0.06,0.55}%gray
\numberwithin{equation}{section}
\begin{document}

%%%%%%%%%%%%%%%%%%%%%%%%%%%%%%%%%%%%%%%%%%%%%%%%
\begin{titlepage}
\begin{flushright}
 \texttt{IFT-UAM/CSIC-15-044}\\
\texttt{KIAS-P15022}\\
\texttt{RIKEN-STAMP-9}\\
\end{flushright}
\vspace{0.5cm}
\begin{center}
{\Large 
\bf  A new 5d description of 6d D-type minimal    conformal matter
}
\\
\vspace{0.9cm}
% Hirotaka Hayashi$^{a}$\footnote{Email: h.hayashi@csic.es},
%  Sung-Soo Kim$^{b}$\footnote{Email: sungsoo.kim@kias.re.kr},
%  Kimyeong Lee$^{b}$\footnote{Email: klee@kias.re.kr},
%  Masato Taki$^{c}$\footnote{Email: taki@riken.jp}\\
% and Futoshi Yagi$^{b}$\footnote{Email: fyagi@kias.re.kr}
Hirotaka Hayashi$^{a}$, 
 Sung-Soo Kim$^{b}$,
 Kimyeong Lee$^{b}$,
 Masato Taki$^{c}$ and Futoshi Yagi$^{b}$\,\footnote{Emails: h.hayashi@csic.es, sungsoo.kim@kias.re.kr, klee@kias.re.kr, taki@riken.jp, fyagi@kias.re.kr}
\\
\vspace{0.3cm}
\it $^a$ Instituto de F\'isica Te\'orica UAM/CSIC, Cantoblanco, 28049 Madrid, Spain
\\
\it $^b$ Korea Institute for Advanced Study (KIAS), \\
85 Hoegiro Dongdaemun-gu, 130-722, Seoul, Korea
\\
\it $^c$  iTHES Research Group, RIKEN, Saitama 351-0198, Japan
\\
\vspace{0.3cm}
\end{center}
\vskip1.0cm
%%%%%%

%%%%%%%%
\begin{abstract}
We propose a new 5d description of  the circle-compactified
  6d $(D_{N+4}, D_{N+4})$ minimal conformal matter  theory which can be approached by   the 6d $\mathcal{N}=(1,0)$ $Sp(N)$ 
gauge theory with $N_f=2N+8$ flavors and one tensor multiplet.  Compactifying the brane set-up for the 6d theory, 
we arrive at a 5-brane Tao diagram for 5d $\mathcal{N}=1$ $SU(N+2)$ theory of the vanishing Chern-Simons level with $2N+8$ flavors.
We conjecture that the 6d theory is recovered as the UV fixed point of this  5d theory. We  show  that the global symmetry of  this 5d theory is  $SO(4N+16)$ identical to that  of the 6d theory   by analyzing the 7-brane monodromy. By using the Tao diagram, we also find the instanton fugacity is exactly given by the circle radius. 
By decoupling flavors in this 5d theory,  one can obtain all the 5d $SU(N+2)$ gauge theories of various Chern-Simons levels and corresponding enhanced global symmetries at the 5d UV fixed point.
\end{abstract}
\end{titlepage}

%%%%%%%%%%%%%%%%%%%%%%%%%%%%%%%%%%%%%%%%%%%%%%%%%%%%%%%%%%%%%%%%%%%%%%

\renewcommand{\thefootnote}{\arabic{footnote}} \setcounter{footnote}{0}

%%%%%%%%%%%%%%%%%%%%%%%%%%%%%%%%%%%%%%%%%%%%%%%%%%%%%%%%%%%%%%%%%%%%%%

\section{Introduction and Conclusion}

Recently there has been an intense investigation of the 6d $\mathcal{N} = (1, 0)$ superconformal field theories (SCFTs) by using the F-theory compactification on elliptically fibered Calabi-Yau spaces and the various brane pictures, which includes
their strings\cite{Haghighat:2013gba,Haghighat:2013tka,Haghighat:2014pva,Kim:2014dza,Haghighat:2014vxa,Gadde:2015tra}, classifications
\cite{Heckman:2013pva,Gaiotto:2014lca,DelZotto:2014hpa,Heckman:2014qba, Heckman:2015bfa, Bhardwaj:2015xxa}, compactifications \cite{Ohmori:2015pua,DelZotto:2015rca}, and 
anomaly polynomials as well as RG flows \cite{Ohmori:2014pca,Ohmori:2014kda,Intriligator:2014eaa,Heckman:2015ola} (also see \cite{DelZotto:2014fia,DelZotto:2015isa,Miao:2015iba}). 
Typical examples are those on the M5 branes near the $E_8$ symmetric M9 wall or on the transverse $ADE$ type singularities or both. 
Interestingly, the 6d SCFT on a single M5-brane probing the $E_8$ symmetric M9-brane \cite{Ganor:1996mu, Seiberg:1996vs} can be also realized as the theory on a single M5-brane probing a $D_4$ singularity, namely the 6d $(D_4,D_4)$ minimal conformal matter \cite{Heckman:2013pva, DelZotto:2014hpa, Ohmori:2015pua}. It is also known that this 6d theory arises as the theory at the UV fixed point of the 5d $SU(2)$ gauge theory with 8 flavors \cite{Seiberg:1996bd, Douglas:1996xp}. Its coupling constant is proportional to the size of the compact circle and the strong coupling limit of the 5d theory would be the 6d theory. By decoupling flavors one by one from the 5d theory, one can realize 5d $SU(2)$ gauge theories with $N_f <8$ flavors which have the 5d strongly coupled fixed point with enhanced $E_{N_f +1}$ global symmetry \cite{Seiberg:1996bd, Morrison:1996xf, Douglas:1996xp}. There are old brane diagram arguments \cite{DeWolfe:1999hj}, recent index calculations \cite{Kim:2012gu, Iqbal:2012xm}, and more recent related arguments to support this symmetry enhancement \cite{Bergman:2013ala, Bao:2013pwa, Hayashi:2013qwa, Taki:2013vka, Bergman:2013aca, Taki:2014pba, Hwang:2014uwa, Zafrir:2014ywa, Hayashi:2014wfa, Mitev:2014isa,Bergman:2014kza, Mitev:2014jza, Kim:2014nqa, Isachenkov:2014eya, Tachikawa:2015mha, Zafrir:2015uaa, Hayashi:2015xla}.

The 5d $SU(2)$ gauge theories have yet another realization as a $(p, q)$ 5-brane web diagram. The $SU(2)$ gauge theories with $0 \leq N_f \leq 7$ flavors have a 
 usual web diagram realization \cite{Aharony:1997ju, Aharony:1997bh,DeWolfe:1999hj} including diagrams suggested in \cite{Benini:2009gi, Kim:2014nqa}.
However, when the number of the flavors reaches eight, we need to consider a new class of the web diagram, so-called Tao web diagram introduced in \cite{Kim:2015jba}, which is given by 5-branes that infinitely expand in a spiral configuration. The appearance of the spiral is interpreted as the $S^1$ for the 6th dimension. An important characteristic of the Tao diagram is that its period of the spiral rotation is identified with the instanton factor of the theory.

From the point of view of the Tao diagram \cite{Kim:2015jba}, the rank of the gauge group is not limited and various Tao diagrams are possible including quiver gauge theories, which may reflect close connection of Tao diagrams to various 6d SCFTs.
However, the relation between the classification of 6d SCFTs and the 5d theories realized by the Tao diagrams was not clear. In this paper, we argue that the $SU(n)$ type Tao diagram can have a definite relation to 6d SCFTs. We identify its 6d origin and claim that it is the generalization of the correspondence between the 5d $SU(2)$ gauge theory with $8$ flavors and the $(D_4, D_4)$ minimal conformal matter on $S^1$. Our main conjecture is that 5d $SU(N+2)$ gauge theory with $N_f = 2N+8$ flavors with the vanishing Chern-Simons (CS) level has the 6d UV fixed point which is the $(D_{N+4}, D_{N+4})$ minimal conformal matter. We also argue that the global symmetry at the fixed point is enhanced to $SO(4N+16)$.

Moreover, starting from the 5d $SU(N+2)$ gauge theory with $N_f = 2N+8$ and decoupling flavors one by one, it is again possible to construct all the 5d $SU(N+2)$ gauge theories with the various CS level which have the 5d UV fixed point. The Tao diagram then reduces to a standard web diagram of $(p, q)$ 5-branes. Bergman and Zafrir have already conjectured the existence of the  UV fixed points for $SU(N+2)$ theories with $N_f\le 2N+7$ in \cite{Bergman:2014kza}.

In the following, we summarize our idea and results. 
In Section \ref{sec:6dto5d}, we consider two paths to support 
our conjecture.
One way is to use the brane set up \cite{Hanany:1997gh, Brunner:1997gf} 
for the 6d $Sp(N)$ gauge theory with $N_f = 2N + 8$ fundamental hypermultiplets and a single tensor multiplet coupled to cancel the gauge anomaly \cite{Seiberg:1996qx, Danielsson:1997kt} by the Green-Schwarz mechanism\footnote{
The 6d theories with gauge group $SU(2)$ and $N_f$ fundamental matter wit a tensor multiplet 
has an additional constraint on $N_f$  by the global anomaly cancellation due to $ \pi_6(SU(2))={\mathbb Z}_{12}$ \cite{Bershadsky:1997sb}. The acceptable number of flavors for the $SU(2)$ gauge case is $N_f=4,10,16$. The 6d counter part for our 5d $SU(3)$ theory with $N_f=10$ is this 6d $Sp(1)$ theory with $N_f=10$.
}, 
which corresponds to the $(D_{N+4}, D_{N+4})$ minimal conformal matter in the tensor branch.
After circle compactification with Wilson line breaking $Sp(N)$ to $U(N)$,
we explain how we obtain 5d $SU(N+2)$ gauge theory with $2N+8$ fundamental hypermultiplets
by brane dynamics.
Another way is to use the description 
as the worldvolume theory on  a single M5-brane probing the orbifold singularity $\mathbb{C}^2/\Gamma_{D_{N+4}}$. 
Compactifying on $S^1$ the M5-brane would yield type IIA setup of a single D4-brane on the same singularity, and the resulting 5d theory is well-known to be associated with the affine D-type Dynkin quiver~\cite{Douglas:1996sw, DelZotto:2014hpa}. 
Here we propose a 5-brane web realization of this quiver theory and the S-dual of this is precisely the $SU(N+2)$ gauge theory with $2N+8$ flavors.
We also show that the expected relation between the gauge couplings and the compactified 
radius is reproduced via Tao diagram.

In Section \ref{sec:globalsymm}, 
we systematically study the global symmetry enhancement of the 5d $SU(n)$ theory with $N_f \le 2n+4$ flavors.%
\footnote{We use the notation $n=N+2$ for convenience.}
We start from the brane web diagram for the 5d $SU(n)$ theory with $2n+4$ flavors and remove the flavor multiplets one by one by making its mass infinite. Each removal generates the half integer Chern-Simons level whose sign depends on the sign of the mass term, leading to a class of 5d $SU(n)_\kappa$ theories with $N_f$ flavors%
\footnote{We use the notation that $SU(n)_\kappa$ represents that $SU(n)$ gauge theory of Chern-Simons level $\kappa$.}
 with constraint $N_f+2|\kappa| \le 2n+4$.
These 5d $SU(n)_{\kappa}$ theories with $N_f$ flavors with constraint $N_f+2|\kappa| \le 2n+4$ have classically $SU(2N_f)\times U(1)_B\times U(1)_I$ global symmetry where $U(1)_B$ is the baryonic part of the flavor symmetry and $U(1)_I$ is the topological symmetry arising from the instanton numbers.  
These theories have an enhanced symmetry at the UV fixed point which is a subgroup of $SO(4n+8)$. 
We calculate enhanced symmetry for all of the $SU(n)_\kappa$ theories with flavors $2n \le N_f\le 2n+4$ and of the $SU(3)_\kappa$ theories with flavors $N_f+2|\kappa| \le 10$.
Especially, we find that the $SU(n)_{\frac12}$ theory with $N_f=2n+3$ flavors has $SO(4n+8)$ global symmetry which is identical to that of the 6d theory. This is a generalization of the 5d $SU(2)$ theory with $N_f = 7$ flavors having an enhanced $E_8$ global symmetry.  
Our result in Section \ref{sec:globalsymm} is consistent with a few cases found recently in \cite{Bergman:2013aca, Bergman:2014kza}, where the authors have studied the enhanced global symmetry for $SU(n)_{n}\ {\rm  with}\ N_f=0$, the most cases of  $SU(3)_\kappa$ with $N_f\le 8$. We enclose all cases for the completeness. 

As summarized above, we provide several evidences for our conjecture in this paper. It would be certainly desirable to find more supports to establish the conjecture. 
 One way is to calculate the index function with very careful treatment of instanton computation from the 5d point of view and show that the global symmetry is enhanced to ${SO}(4N+16)$ \cite{WorkInProgress}. A similar calculation has been done for the $SU(2)$ case with $E_8$ global symmetry \cite{Hwang:2014uwa, Kim:2014dza}. As done in \cite{Kim:2015jba},  it is possible to compute the partition function applying the topological vertex method to the Tao diagram  \cite{WorkInProgress}. 
Another way is to start from the 6d point of view. With the introduction of the $Sp(N)$ Wilson line, together with the contribution from the tensor multiplet, the elliptic genus calculation of instanton strings may lead to the 5d $SU(N+2)$ gauge symmetry in the Coulomb phase \cite{WorkInProgress}. 
Our observation for the relation between  the 6d and 5d supersymmetric theories may lead to further examples and insights.

\paragraph{Note added:}
We here note that our results have some overlap with \cite{Yonekura:2015ksa}, which appeared in arXiv on the same day.

%%%%%%%%%%%%%%%%%%%%%%%%%%%%%%%%%%%%%%%%%%%%%%%%%%%
\section{From 6d to 5d}\label{sec:6dto5d}

In this section, we justify our main conjecture from two different routes. Our main conjecture is that the 5d $SU(N+2)$ gauge theory with $N_f = 2N+8$ flavors and the vanishing CS level has the 6d UV fixed point that is the $(D_{N+4}, D_{N+4})$ minimal conformal matter. In other words, a circle compactification of the $(D_{N+4}, D_{N+4})$ minimal conformal matter with Wilson line along it gives the the 5d $SU(N+2)$ gauge theory with $N_f = 2N+8$ flavors and the zero CS level. Hence, our approach is to start from the D-type minimal conformal matter on $S^1$ and then arrive at the 5d $SU(N+2)$ gauge theory with $N_f = 2N+8$ flavors and the zero CS level. 

\begin{figure}
\begin{center}
\includegraphics[width=15cm]{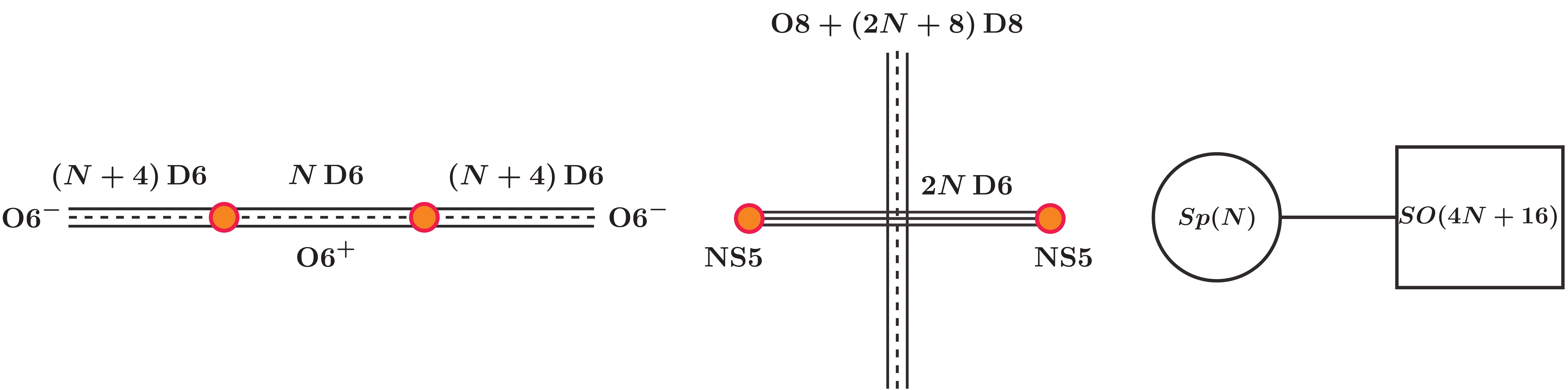}
 \end{center}
\caption{Left: Type IIA brane realization of the $(D_{N+4}, D_{N+4})$ minimal conformal matter in the tensor branch. Middle: Another brane realization of the same 6d theory. Right: The quiver diagram of the 6d theory. } 
\label{Fig:brane}
\end{figure}

In section \ref{sec:evidence1}, we first go to the tensor branch of the $(D_{N+4}, D_{N+4})$ minimal conformal matter. An $S^1$ compactification of the system is given by $N$ D6-branes on top of an O6$^+$-plane suspended between two NS5-branes and also semi-infinite $N+4$ D6-banes on top of an O6$^-$-plane on the left-hand side of the left NS5-brane and the right-hand side of the right NS5-brane in type IIA string theory \cite{DelZotto:2014hpa,Hanany:1997gh, Brunner:1997gf}. 
The configuration is depicted in the leftmost diagram in Figure \ref{Fig:brane}.  
%%%%%%%%%%%%%%%%%%%%%
%%%%%%%%%%%%%%%%%%%%%
The resulting theory is the 6d $Sp(N)$ gauge theory with $N_f = 2N+8$ flavors and a tensor multiplet. In fact, the same 6d theory can be also realized by another brane set up in type IIA string theory depicted in the middle diagram in Figure \ref{Fig:brane}. In this case, we have again $2N$ D6-branes%
\footnote{The $2N$ includes the mirror image of D6-branes .}
suspended between NS5-brane and its mirror image through $O8^-$.
However, in order to realize the $Sp(N)$ gauge group and also flavors, we introduce $2N+8$ D8-branes on top of an O8$^-$-plane \cite{Hanany:1997gh, Brunner:1997gf}. Then, we again obtain the 6d $Sp(N)$ gauge theory with $N_f = 2N+8$ flavors and a tensor multiplet coupled. The quiver diagram (without a tensor multiplet) of the 6d theory is depicted in the rightmost diagram in Figure \ref{Fig:brane}. This type IIA brane set up is the starting point in section \ref{sec:evidence1}. 

In section \ref{sec:evidence2}, we use a different route. We first reduce the 6d $(D_{N+4}, D_{N+4})$ minimal conformal matter on $S^1$ with the Wilson line along it, and go to the 5d affine $\hat{D}_{N+4}$ quiver theory. We propose a $(p, q)$ 5-brane web diagram which realizes the 5d affine $\hat{D}_{N+4}$ quiver theory, and the IIB brane set up is the starting point in section \ref{sec:evidence2}.

\subsection{Evidence 1: O-plane description}
\label{sec:evidence1}
We now explain the main conjecture by using the brane description depicted in the middle diagram in Figure \ref{Fig:brane}. As discussed before, this 6d $Sp(N)$ theory is known to have the type IIA brane description
with $N$ D6 branes suspended between two NS5-branes,
with the insertion of $2N+8$ D8 branes and one O8$^-$-plane \cite{Hanany:1997gh, Brunner:1997gf}.
We compactify one of the direction parallel to all these branes
with the Wilson loop. 
Taking T-dual along this compactified direction,
we obtain the type IIB brane setup with $N$ D5 branes, two NS5 branes, 
$2N+8$ D7 branes and two O7$^-$-planes.
Due to the Wilson loop,
the $N$ D5 branes appear away from both of the two O7$^-$-planes,
breaking $Sp(N)$ gauge group to $U(N)$.

\begin{figure}[t]
\begin{center}
\includegraphics[width=6cm]{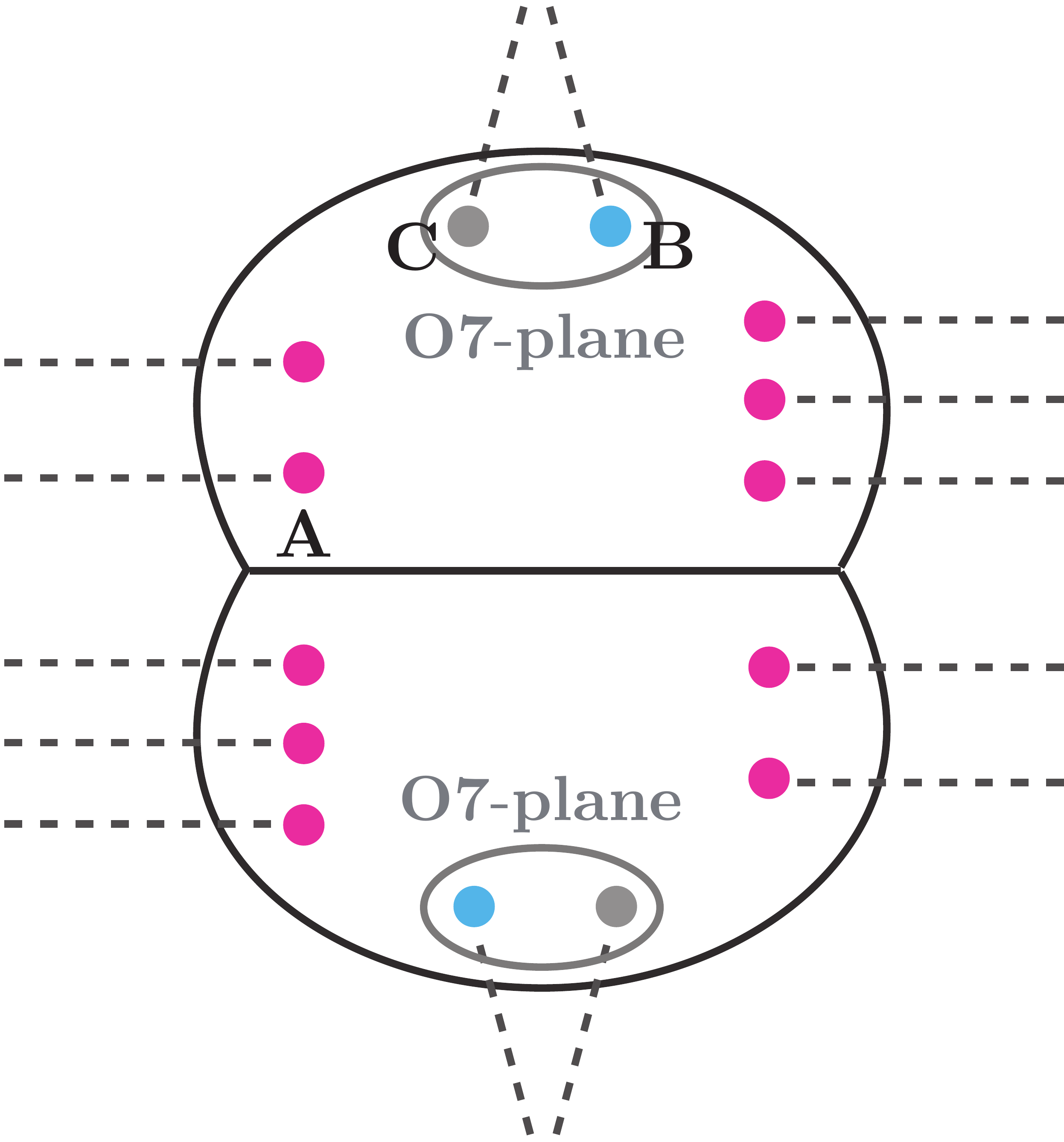}
 \end{center}
\caption{The T-dual description of the type IIA  brane setup for 6d $\mathcal{N}=(1,0)$ $Sp(N)$ gauge theory
with $2N+8$ flavors and one tensor multiplet. (For simplicity, the brane description for the $N=1$ case is drawn.) We denote $\mathbf{A}$, $\mathbf{B}$, and $\mathbf{C}$ for $[1,0]$ 7-brane (or
D7 brane), $[1,-1]$ 7-brane, and $[1,1]$ 7-brane, respectively.
The two $O7$-planes are replaced by 
the pair of $\mathbf{B}$ and $\mathbf{C}$. The branch cuts due to 7-branes are denoted by the dotted lines.} 
\label{Fig:7in}
\end{figure}

\begin{figure}[t]
\begin{center}
\includegraphics[width=8cm]{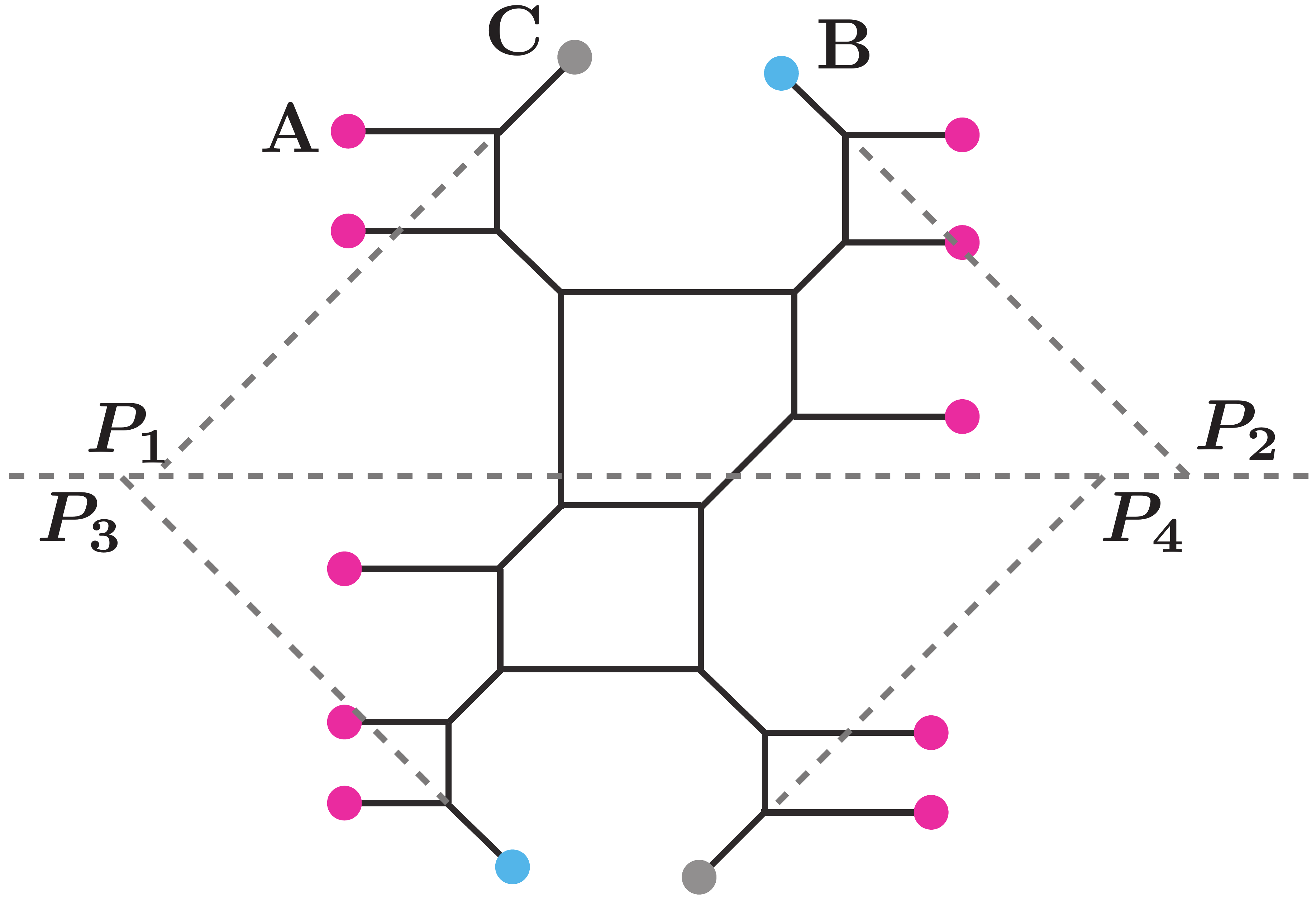}
 \end{center}
\caption{A $(p,q)$-brane setup for 5d $SU(N+2)$ gauge theory with $2N+8$ flavors. For concreteness, the $N=1$ case is drawn. The horizontal dotted line indicates the center of mass position of $N+2$ color branes. To measure the gauge coupling, one extends the upper and lower $(1,1)$ and $(1,-1)$ 5-branes to the horizontal dotted line, which gives rise to two asymptotic distances between $\mathbf{P_1}$ and $\mathbf{P_2}$, and between $\mathbf{P_3}$ and $\mathbf{P_4}$.
The inverse gauge coupling is then given by the average of these 
two distances.}
\label{Fig:pq}
\end{figure}

We note that quantum resolution of O7$^-$-plane is given by the pair of 
$[1,1]$ 7-brane and $[1,-1]$ 7-brane \cite{Sen:1996vd}.
When we apply this property to the two O7$^-$-planes in this brane setup,
we assume that a ``5-brane loop'' \cite{DeWolfe:1999hj} appears,
where the original two NS-branes being part of this 5-brane loop 
and all the 7-branes are inside the 5-brane loop. 
The $Sp(1)$ case is depicted in Figure \ref{Fig:7in}. 
Note that the 5-brane charge between 
the cut created by the $[1,1]$ 7-brane and the $[1,-1]$ 7-brane should be $(1,0)$,
implying that, as a consequence, two more color D5-branes are generated in this process\footnote{An analogous process is discussed, for example, in \cite{DeWolfe:1999hj} 
relating the type IIA brane setup with an O8-plane for the 5d $Sp(1)$ gauge theory
to the type IIB $(p,q)$ brane web description. Contrary to our case,
two D5-branes become a part of the 5-brane loop and two more NS5-branes are generated.}.
Therefore, we obtain $N+2$ D5-branes in total.
When we pull out the 7-branes outside the 5-brane loop
by taking into account the Hanany-Witten transition \cite{Hanany:1996ie},
we obtain the $(p, q)$ 5-brane web, Figure \ref{Fig:pq}, where the uppermost and lowermost D5 branes  are created.
In this diagram, we can explicitly see that there are $N+2$ color D5-branes 
indicating the $SU(N+2)$ gauge symmetry.
Therefore, we expect that the gauge symmetry is enhanced from $U(N)$ to $SU(N+2)$.

In our case, replacing the two O7$^-$-planes by two pairs of $[1,1]$ and $[1,-1]$ 7-branes has another important aspect.
%Before this replacement, we see that the compactified direction is $S^1/\mathbb{Z}_2$ due to the two O7$^-$-planes where the size of $S^1$ is the separation between two O7$^-$-planes.
Before this replacement, two O7$^-$-planes are located at two places in the compactified circle such that the distance between two O7$^-$-planes is the half circumference of the circle.
%where the size of $S^1$ is the separation between two O7$^-$-planes.
With the replacement, we are changing the picture so that we look at the theory from the purely 5d point of view. Hence,
%After the replacement, 
the notion of the compactified direction is transmuted into the gauge coupling of the theory.
%this compactified direction is no longer compact. 
We can see that the 
radius of the compactified circle %$S^1/\mathbb{Z}_2$ 
is now inherited by the the inverse gauge coupling in the $(p, q)$ 5-brane web diagram,
which is identified as the average of the asymptotic distances
between $(1,1)$ 5-brane and $(1,-1)$ 5-brane measured 
at the center of mass position  of $N+2$ color D5-branes by extrapolating them as in Figure \ref{Fig:pq}
(up to a numerical factor depending on the convention).
Therefore, we expect that the KK mode of the original 6d theory compactified on $S^1$ is reinterpreted as 
the instanton contribution in the 5d theory. 
\subsection{\texorpdfstring{
Evidence 2: $\hat{D}$-type Dynkin quiver}
{Evidence 2: \hat D-type Dynkin quiver}
}
\label{sec:evidence2}

\begin{figure}[t]
\begin{center}
\includegraphics[width=8cm]{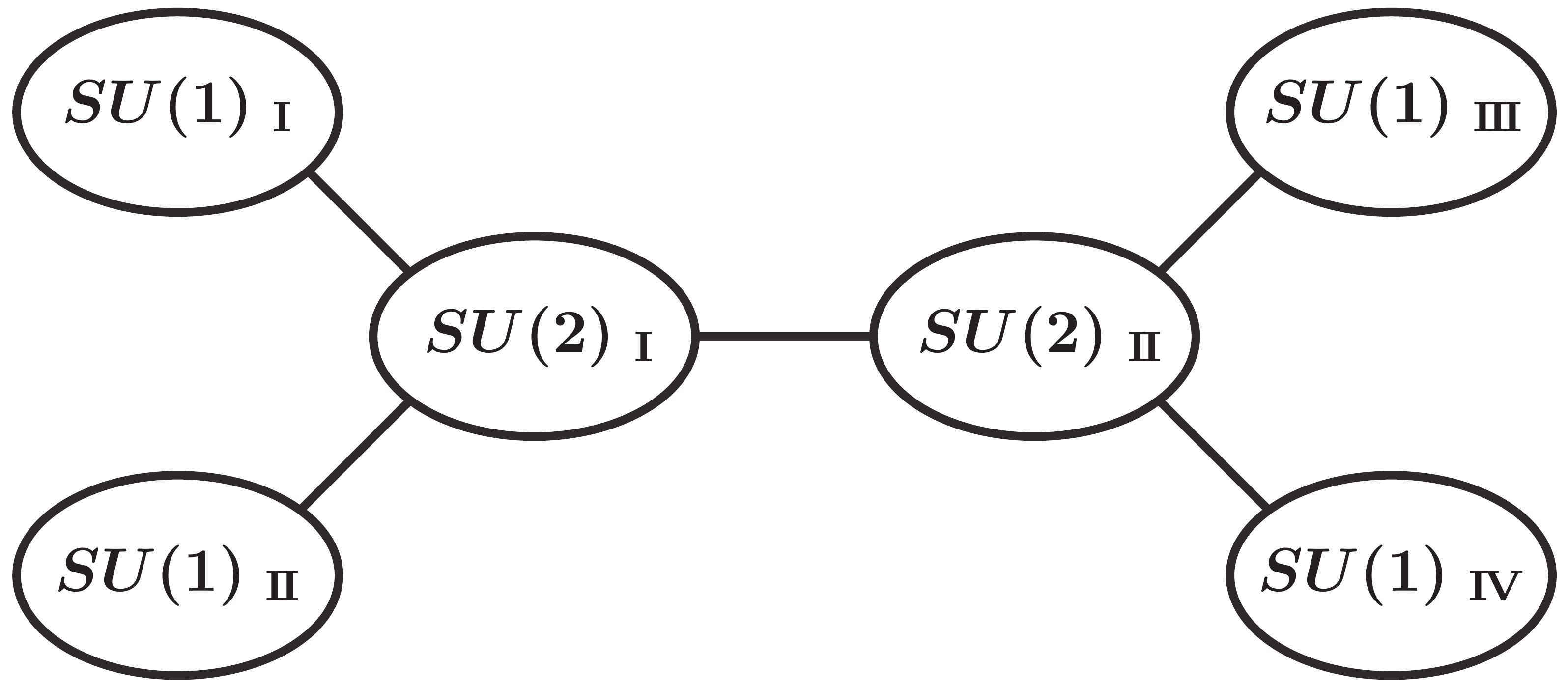}
 \end{center}
\caption{The quiver diagram of the affine $\hat{D}_5$ quiver theory.} 
\label{Fig:D5dynkin}
\end{figure}

It is also possible to justify our conjecture from a different route. 
The 6d $(D_{N+4}, D_{N+4})$ minimal conformal matter is realized on a single M5-brane at the orbifold singularity of $\mathbb{C}^2/\Gamma_{D_{N+4}}$ \cite{DelZotto:2014hpa}.
An $S^1$ compactification of the M5-brane may yield a system of a single D4-brane sitting at the same singularity of $\mathbb{C}^2/\Gamma_{D_{N+4}}$ in type IIA string theory. The resulting five-dimensional theory is a 5d affine $\hat{D}_{N+4}$ quiver theory. The quiver diagram of the case $N=1$ is depicted in Figure \ref{Fig:D5dynkin}. 
%%%%%%%%%%%%%%%%%%%%%%%%%%%%%%%
%%%%%%%%%%%%%%%%%%%%%%%%%%%%%%%
In the quiver diagram, the $N+1$ nodes along the central horizontal line give $SU(2)$ gauge groups, and the $4$ nodes on the left-hand and right-hand side of the diagram give ``$SU(1)$" gauge groups. Each line which connects two gauge nodes represents a bi-fundamental hypermultiplet. Note that 
$SO(2N+8)\times SO(2N+8)$ global symmetry of the six-dimensional theory is broken to a subgroup\footnote{However, we will see that there is an enhancement of the global symmetry.}
due to Wilson line along the $S^1$.

Let us comment on the meaning of the ``$SU(1)$" gauge groups. In fact a set of a bi-fundamental hypermultiplet of $SU(2) \times SU(1)$ and an ``$SU(1)$" gauge instanton is equivalent to two fundamental hypermultiplets which are coupled to the $SU(2)$ gauge group. This has been confirmed in various examples at least at the level of the 5d partition functions in \cite{Hayashi:2013qwa, Bao:2013pwa, Bergman:2014kza, Hayashi:2014hfa}. Hence, the 5d affine $\hat{D}_{N+4}$ quiver theory may be equivalent to a 5d $SU(2)^{N+1}$ linear quiver gauge theory with four fundamental hypermultiplets are coupled to the two $SU(2)$ gauge group at the left and right ends, respectively.
However, in the following discussion, we keep the convention of ``$SU(1)$'' gauge group.

%%%%%%%%%%%%%%%%%%%%%%%%%%%%%%%
\begin{figure}[t]
\begin{center}
\includegraphics[width=7cm]{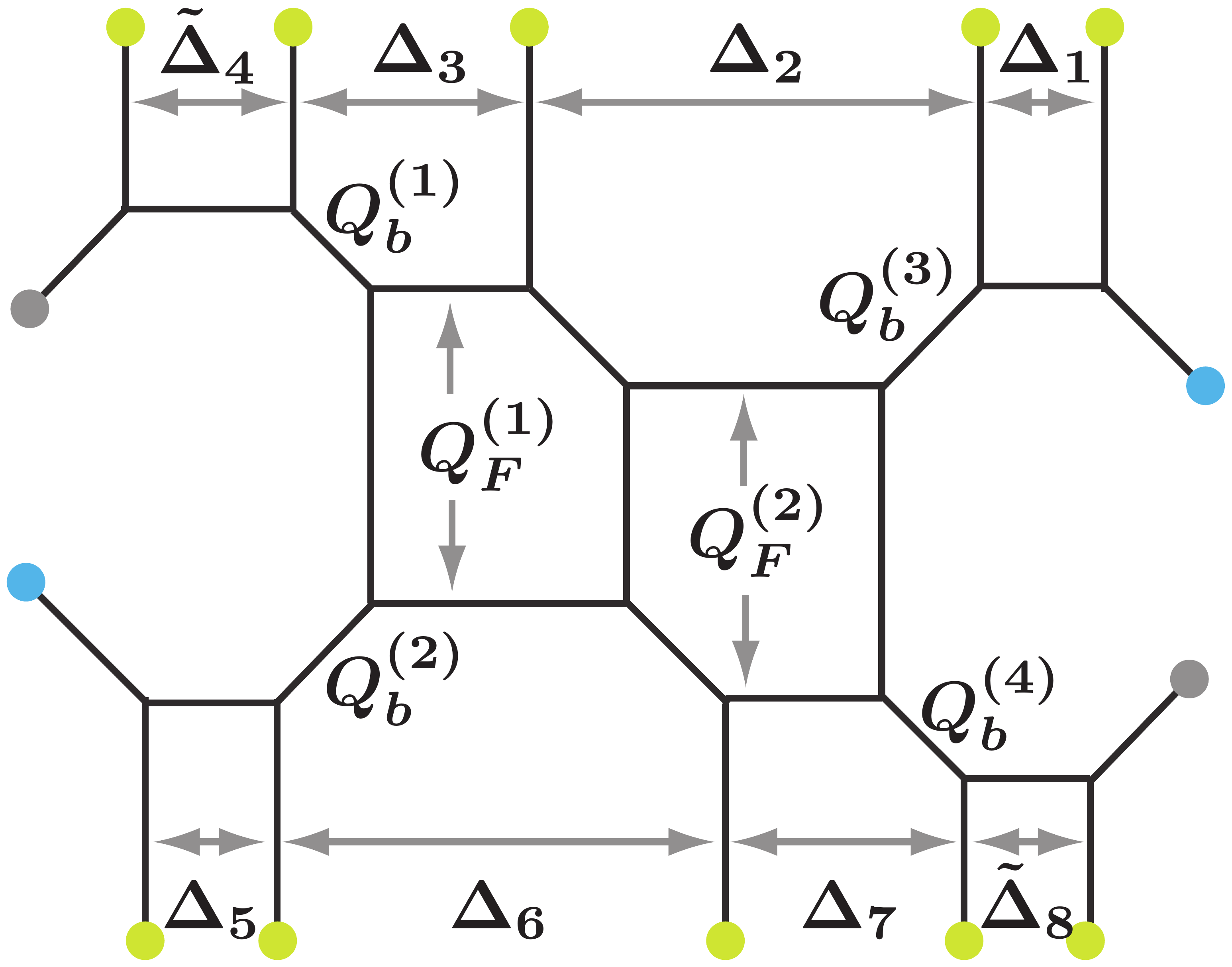}
 \end{center}
\caption{The web diagram which realizes an affine $\hat{D}_5$ Dynkin quiver theory. Q and $\Delta$ assigned to a 5-brane
segment in the figure represent $e^{-L}$ where $L$ is the length of the corresponding 5-brane. This can be also seen as a diagram S-dual to the web diagram for $SU(3)$ gauge theory with $10$ flavors.}
\label{Fig:Sp1-Sp1}
\end{figure}
%%%%%%%%%%%%%%%%%%%%%%%%%%%%%%%

We propose that this 5d affine $\hat{D}_{N+4}$ quiver theory can be also realized by a web diagram of $(p, q)$ 5-branes in type IIB string theory. 
The web diagram that we propose in the case with $N=1$ is depicted in Figure \ref{Fig:Sp1-Sp1}. 
In this web diagram, the two connected closed faces in the middle of the web give $SU(2) \times SU(2)$ gauge theory which 
%, and the two gauge theories 
is coupled by a bi-fundamental hypermultiplet. % of the $SU(2) \times SU(2)$ gauge groups. 
For the sake of the later computation, let the gauge group from the left loop be $SU(2)_I$ and the gauge group from the right loop be $SU(2)_{II}$. Furthermore, we have four ``$SU(1)$" gauge groups corresponding to four horizontal internal lines with lengths $\Delta_1, \tilde{\Delta}_4, \Delta_5$ and $\tilde{\Delta}_8$ in Figure \ref{Fig:Sp1-Sp1}. Each ``SU(1)" is coupled to either of the two $SU(2)$ gauge groups by a bi-fundamental hypermultiplet. We will denote the four ``$SU(1)$" gauge groups coming from the horizontal lines with lengths $\tilde{\Delta}_4, \Delta_5, \Delta_1$ and $\tilde{\Delta}_8$ by  $SU(1)_I, SU(1)_{II}, SU(1)_{III}$ and $SU(1)_{IV}$ respectively. Therefore, we claim that the web diagram in Figure \ref{Fig:Sp1-Sp1} exactly yields the affine $\hat{D}_5$ Dynkin quiver theory in Figure \ref{Fig:D5dynkin}.  It is straightforward to write down a web diagram for general $N$. Namely, we put more closed faces along the horizontal directions.

One can give a completely different view by the S-duality rotating the web diagram by $90$ degrees. After the S-duality, the web diagram which originally yielded the affine $\hat{D}_{N+4}$  quiver theory becomes the web diagram which realizes the $SU(N+2)$ gauge theory with $2N+8$ fundamental hypermultiplets.  For the $N=1$ case, one can easily see that the web diagram in Figure \ref{Fig:Sp1-Sp1} is nothing but the web diagram in Figure \ref{Fig:pq} after the S-duality.  Therefore, we again obtain the $SU(N+2)$ gauge theory with $2N+8$ flavors from the $S^1$ compactification of the six-dimensional theory on a single M5-brane at the origin of $\mathbb{C}^2/\Gamma_{D_{N+4}}$.

From this point of view of the affine $\hat{D}_{N+4}$ Dynkin quiver theory, one can further give a quantitative support for our conjecture. For simplicity, let us focus on the case with $N=1$. Since the $S^1$ compactification gives the 5d affine $\hat{D}_5$ quiver theory, an appropriate product of the instanton fugacities of the quiver theory is expected to be equal to the fugacity for the KK $U(1)$ symmetry. The exponents of the instanton fugacity of a gauge node are the comarks.
The exponent for instanton fugacity associated to the extended node is 1, for instance. More specifically, the following relation should hold for the affine $\hat{D}_5$ quiver, 
\begin{equation}
q^{SU(1)_I}q^{SU(1)_{II}}\left(q^{SU(2)_{I}}\right)^2\left(q^{SU(2)_{II}}\right)^2q^{SU(1)_{III}}q^{SU(1)_{IV}}=\Delta,
\label{period}
\end{equation}
where $q^{G}$ is the instanton fugacity for a gauge group $G$, and $\Delta$ is the fugacity for the KK $U(1)$ symmetry. 
In the following, we discuss that the relation \eqref{period} is indeed satisfied from the web diagram in Figure \ref{Fig:Sp1-Sp1}.

\begin{figure}
\begin{center}
\includegraphics[width=13cm]{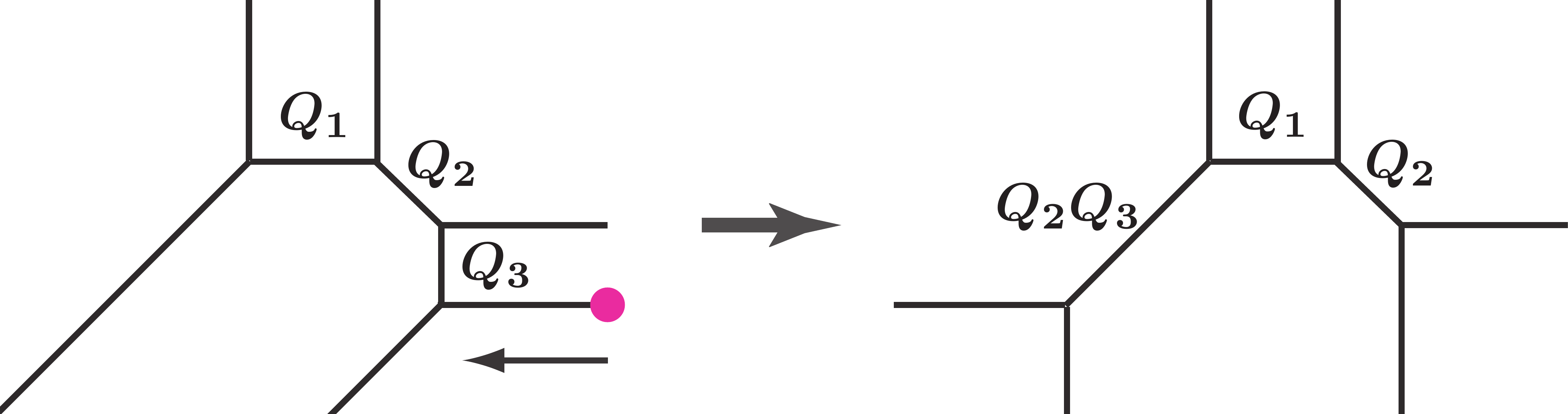}
 \end{center}
\caption{Left: the web diagram which gives an ``SU(1)" gauge theory. Right: another equivalent web diagram which is given by moving the 7-brane in the left figure. This figure makes it easy to identify the instanton fugacity for the ``$SU(1)$". } 
\label{Fig:SU(1)}
\end{figure}

The instanton fugacity of each gauge group can be easily inferred from the web diagram. Namely it is given by an exponential of the average of the sum of the asymptotic distances between two NS5-branes extending in the upper and the lower part of diagram (see for example \cite{Bao:2011rc}). Hence, the instanton fugacities for the two $SU(2)$ gauge groups are
\begin{equation}
q^{SU(2)_{I}} = \left(\Delta_3\Delta_6\right)^{\frac{1}{2}}, \qquad q^{SU(2)_{II}} = \left(\Delta_2\Delta_7\right)^{\frac{1}{2}}.
\label{qsu2}
\end{equation}
The instanton fugacities of the ``SU(1)" gauge groups can be understood in a similar way. Let us consider the diagram in Figure \ref{Fig:SU(1)}.
%%%%%%%%%%%%%%%%%%%%%%%%%%%%%%%
%%%%%%%%%%%%%%%%%%%%%%%%%%%%%%%
The identification of the instanton fugacity of the ``$SU(1)$" gauge group is less obvious in the left diagram in Figure \ref{Fig:SU(1)}. However, one can move the 7-brane, and after the Hanany-Witten transition one arrives at the diagram on the right hand side of Figure \ref{Fig:SU(1)} 
which gives an equivalent theory. From the diagram on the right, 
it is then easy to identify the asymptotic distances between two NS5 branes: the upper and lower distances are $Q_1$ and $Q_1Q_2^2Q_3$, respectively. 
The instanton fugacity is thus given by 
\begin{equation}
q^{SU(1)} %= \Big[Q_1 \cdot (Q_1Q_2^2Q_3) \Big]^\frac12 
= Q_1Q_2Q_3^{\frac{1}{2}}.  \label{SU(1)}
\end{equation}
One can apply the rule \eqref{SU(1)} to the instanton fugacity of each of the four ``$SU(1)$" gauge groups in Figure \ref{Fig:Sp1-Sp1}, and the result is 
\begin{equation}\begin{split}
&q^{SU(1)_I} = \tilde{\Delta}_4Q_b^{(1)}Q_F^{(1)}{}^{\frac{1}{2}}, \qquad q^{SU(1)_{II}} = \Delta_5Q_b^{(2)}Q_F^{(1)}{}^{\frac{1}{2}}, \\
& q^{SU(1)_{III}} = \Delta_1Q_b^{(3)}Q_F^{(2)}{}^{\frac{1}{2}}, \qquad q^{SU(1)_{IV}} = \tilde{\Delta}_8Q_b^{(4)}Q_F^{(2)}{}^{\frac{1}{2}}.
\label{qsu1}
\end{split}\end{equation}

\begin{figure}
\begin{center}
\includegraphics[width=12cm]{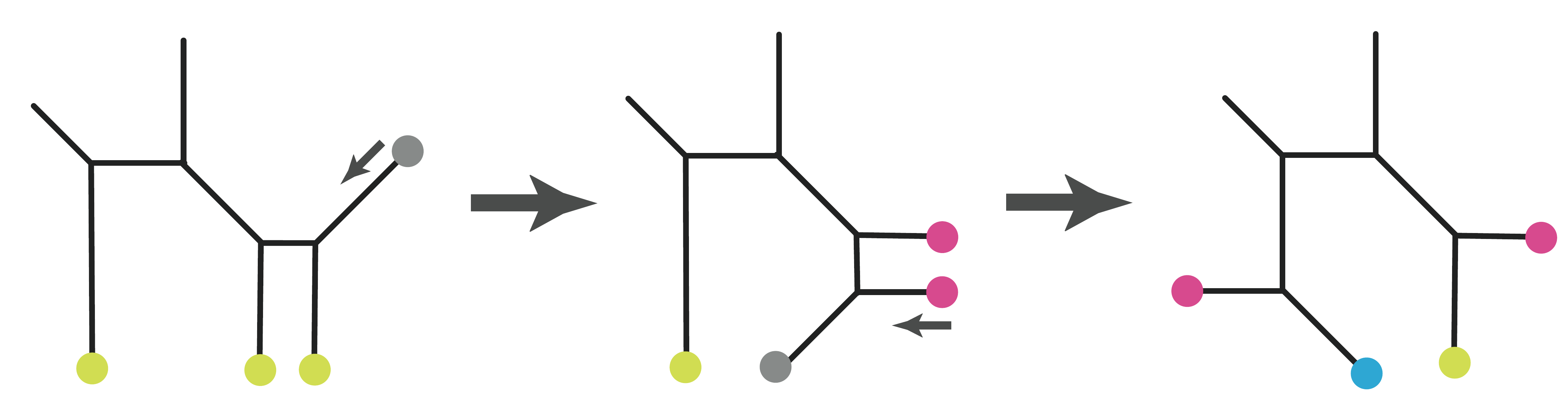}
 \end{center}
\caption{An example of successive Hanany-Witten transition applied on Figure \ref{Fig:Sp1-Sp1} which makes a Tao configuration} 
\label{Fig:intermediate}
\end{figure}

\begin{figure}
\begin{center}
\includegraphics[width=12cm]{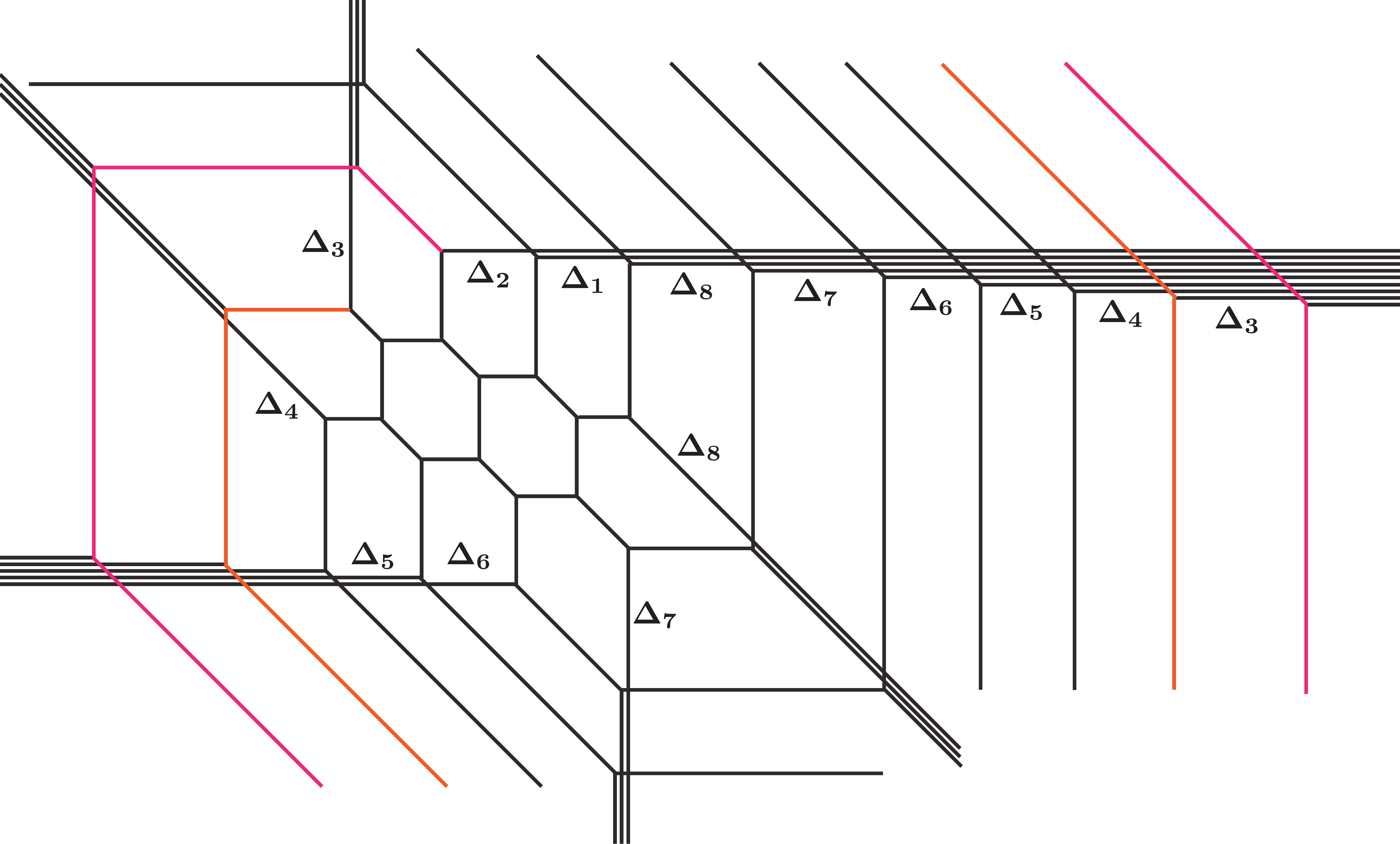}
 \end{center}
\caption{The Tao web diagram corresponding to the affine $\hat{D}_5$ Dynkin diagram in Figure \ref{Fig:Sp1-Sp1}.} 
\label{Fig:Tao}
\end{figure}

To relate them with the KK $U(1)$ fugacity $\Delta$, one needs to move to the Tao diagram from Figure \ref{Fig:Sp1-Sp1} by using a sequence of Hanany-Witten transition 
and moving out all the 7-branes to infinity as discussed in \cite{Kim:2015jba}. 
For example, from Figure \ref{Fig:Sp1-Sp1}, one can perform a flop transition associated with $Q^{(2)}_{b}$ as well as $Q^{(3)}_b$, and then do successive Hanany-Witten transition on the right bottom and on the left top 7-branes involving % $\mathbf{C}$ 
$[1,1]$ 7-brane (gray dot) and three $[0,1]$ 7-branes % $\mathbf{N}$ 
(dark yellow dots) as in Figure \ref{Fig:intermediate}.
\begin{figure}
\begin{center}
\includegraphics[width=14cm]{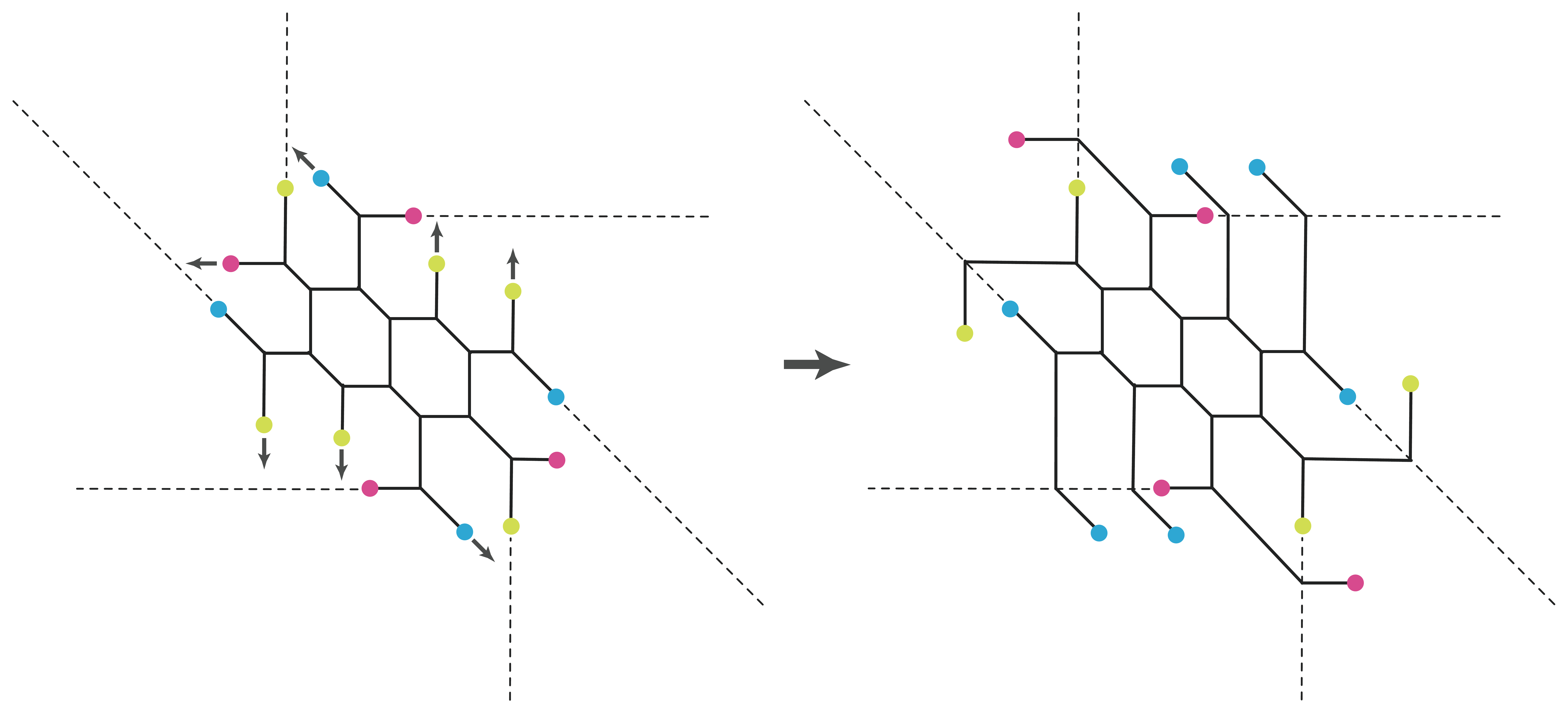}
 \end{center}
\caption{The 7-brane motions leading to Tao web diagram for the affine $\hat{D}_5$ Dynkin diagram .} 
\label{Fig:su2xsu2}
\end{figure}
After choosing proper branch cuts as in Figure \ref{Fig:su2xsu2}, one takes all the 7-branes to infinity to obtain a Tao diagram depicted in Figure \ref{Fig:Tao}.
%%%%%%%%%%%%%%%%%%%%%%%%%%%%%%%
%%%%%%%%%%%%%%%%%%%%%%%%%%%%%%%
In this web diagram description, the fugacity $\Delta$ is given by the period of the spiral diagram \cite{Kim:2015jba}. 
The period of the Tao diagram in Figure \ref{Fig:Tao} is simply given by
\begin{equation}
\Delta = \Delta_1\Delta_2\Delta_3\Delta_4\Delta_5\Delta_6\Delta_7\Delta_8. \label{rhs.period}
\end{equation}
From Figure \ref{Fig:Tao}, it is also possible to identify $\Delta_4, \Delta_8$ by the parameters in Figure \ref{Fig:Sp1-Sp1}, and the result is
\begin{equation}
\Delta_4 = \tilde{\Delta}_4Q_b^{(1)}Q_b^{(2)}Q_F^{(1)}, \qquad \Delta_8 = \tilde{\Delta}_8Q_b^{(3)}Q_b^{(4)}Q_F^{(2)}.  \label{relation}
\end{equation}
Using \eqref{qsu2}, \eqref{qsu1} and \eqref{relation}, one can rewrite the left hand side of \eqref{period} as 
\begin{equation}
q^{SU(1)_I}q^{SU(1)_{II}}\left(q^{SU(2)_I}\right)^2\left(q^{SU(2)_{II}}\right)^2q^{SU(1)_{III}}q^{SU(1)_{IV}}=  \Delta_1\Delta_2\Delta_3\Delta_4\Delta_5\Delta_6\Delta_7\Delta_8,
\end{equation}
which is exactly equal to $\Delta$ due to \eqref{rhs.period}. Hence, the relation \eqref{period} indeed holds for the affine $\hat{D}_5$ Dynkin web diagram in Figure \ref{Fig:Sp1-Sp1}. The same argument in fact holds for the $SU(3)$ gauge theory with $10$ flavors. 
The generalization of the computation to general  $N$ is also straightforward.

%%%%%%%%%%%%%%%%%%%%%%%%%%%%%%%%%%%%%%%%%%%%%%%%%%%
\section{\texorpdfstring{
Global symmetry of $\boldsymbol{SU(n)}$ gauge theories}{Global symmetry of SU(n) gauge theories}}\label{sec:globalsymm}

\begin{figure}
 \begin{center}
\includegraphics[width=13cm]{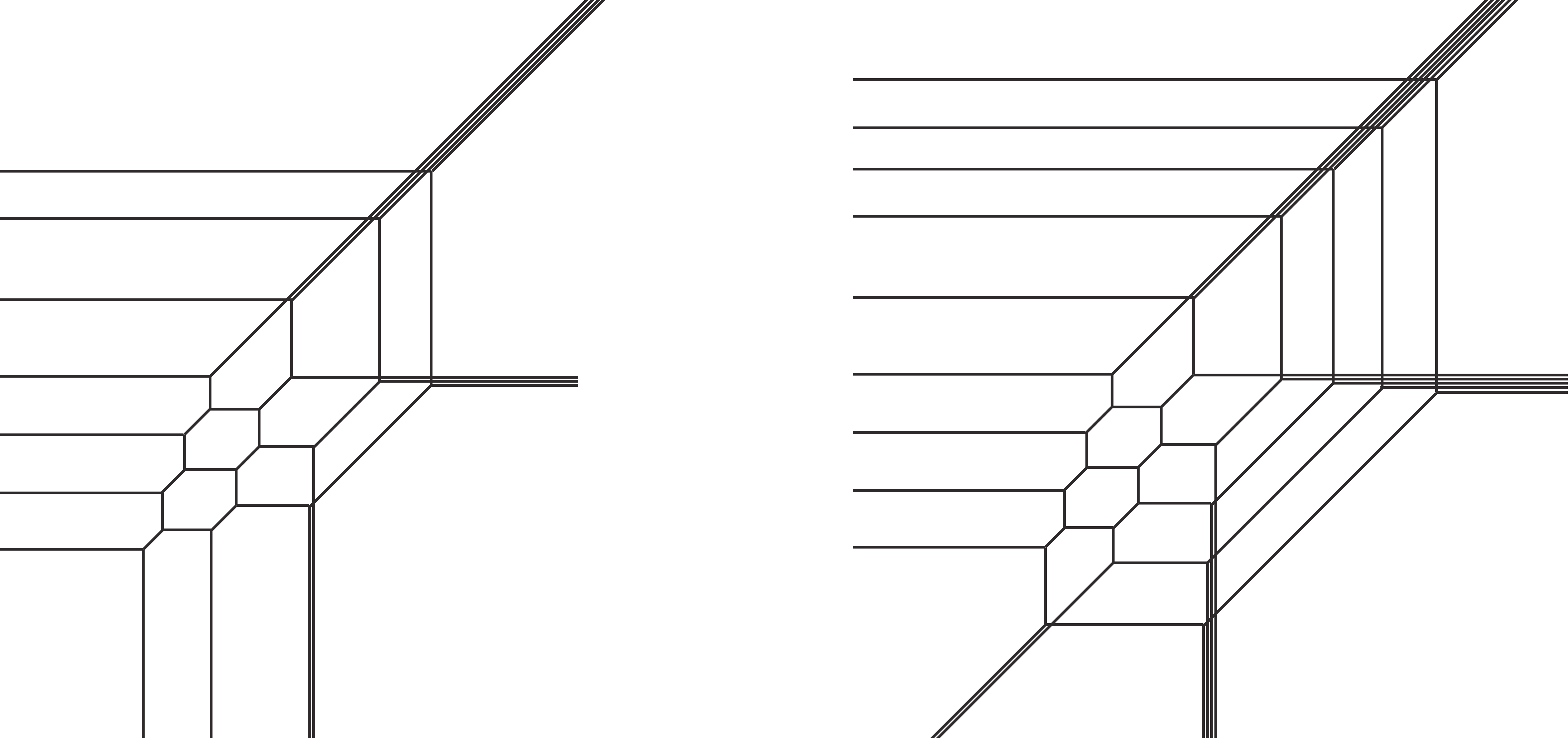}
 \end{center}
 \caption{A web diagram of $SU(3)$ theory with $N_f=8$ flavors with CS level 1 (left) and with $N_f=9$ flavors with CS level $\frac12$ (right). They are obtained by taking mass decoupling limit from Tao diagram.}
 \label{fig:nf89}
\end{figure}

The classification of the UV complete 5d $SU(n)$ theories\footnote{Throughout this section, instead of  $SU(N+2)$ gauge theories used in the previous section, we use $SU(n)$ for denoting the 5d gauge group for simplicity. To relate to Section \ref{sec:6dto5d}, one can take $n=N+2$.
  } with $n > 2$ and $N_f$ fundamental hypermultiplets has been done in \cite{Intriligator:1997pq} which states that 5d UV fixed point exists for $N_f\leq 2n$. 
On the other hand, it was conjectured in \cite{Bergman:2014kza} that the 5d UV fixed point exists for
\begin{align}\label{5dfixedptcond}
 N_f < 2n+4, \qquad  {\rm and}\qquad N_f\leq 2n+4- 2|\kappa|,
\end{align}
 where $\kappa$ is the Chern-Simons (CS) level.  For $n=3$, the  $(p,q)$ 5-brane web for $N_f\le 6$ is straightforward. 
The web diagrams for $N_f= 7$ with $\kappa= \frac12, \frac32$ and $N_f=8$ with $\kappa=0$ are given in \cite{Bergman:2014kza}. 
 Here we   strengthen this conjecture by finding  the additional diagrams for $N_f=9$ with $\kappa=\frac12$ and $N_f=8$ with $\kappa=1$ in Figure \ref{fig:nf89}.

 The insight of \cite{Kim:2015jba} is that one can extend the class of the web diagrams by including Tao diagrams where 5-branes has an infinitely expanding spiral configuration, and the corresponding 5d theories are expected to be UV complete and have a 6d fixed point.  From this point of view, our conjecture is that the UV fixed point of the 5d $SU(n)$ gauge theory which satisfies the bound $N_f = 2n+4$ exists due to the existence of the corresponding Tao diagram\footnote{For example, the Tao diagram of $n=3$ is given by Figure \ref{Fig:Tao}.},  and this UV SCFT is precisely the 6d $(D_{n+2}, D_{n+2})$ minimal conformal matter. 
This is essentially a generalization of the claim for the known $n=2$ case.

In the $n=2$ case, the global symmetry is enhanced either at the 5d fixed point or the 6d fixed point depending on the UV completion. In this section, we scrutinize the global symmetry enhancement for 5d $SU(n)$  gauge theory with $n> 2$ and $2n\le N_f\le 2n+4$ flavors together with the CS level $\kappa$ subject to \eqref{5dfixedptcond} We also list up the non-abelian part of the global symmetries of all the possible $SU(3)$ gauge theories with flavors and various CS levels which have the UV fixed point. We employ the 7-brane technology developed by \cite{Gaberdiel:1997ud, Gaberdiel:1998mv, DeWolfe:1999hj} to characterize the global symmetry structure for various flavors.

Given two 7-branes of whose configuration is denoted by $\mathbf{X}_{\boldsymbol{z}_1}$ and $\mathbf{X}_{\boldsymbol{z}_2}$ with two charge vectors $\boldsymbol{z}_i = [{p}_i,{q}_i]$, the changes in the charge of one 7-brane due to crossing the branch cut of the adjoining 7-brane are given by
\begin{align}\label{7branemono}
{\mathbf{X}_{\boldsymbol{z}_1}}
 {\mathbf{X}_{\boldsymbol{z}_2}} 
 &= \mathbf{X}_{\boldsymbol{z}_2}\, 
 \mathbf{X}_{\boldsymbol{z}_1+
 (\boldsymbol{z}_1\wedge \boldsymbol{z}_2)\boldsymbol{z}_2}
=
 \mathbf{X}_{\boldsymbol{z}_2+
 (\boldsymbol{z}_1\wedge \boldsymbol{z}_2)\boldsymbol{z}_1}
{\mathbf{X}_{\boldsymbol{z}_1}},
\end{align}
 where $\boldsymbol{z}_1\wedge \boldsymbol{z}_2=
 {p}_1 {q}_2- {p}_2 {q}_1$. 
As we will see below, a 5-brane web can be recast into 7-brane configuration with 5-brane loop probing it.
The 7-brane configuration carries the enhanced global symmetry,
and 
one can identify the symmetry by
extracting
the 7-brane configurations which collapse into $\mathbf{A}_m, \mathbf{D}_m, \mathbf{E}_m$ Kodaira singularities of which realize $A_m, D_m, E_m$ algebras. 
Rearranging 7-branes with the above reordering rule enables us to extract such basic 7-brane configurations.
If we use a short hand notation for frequently appearing 7-branes: 
\begin{align}
\mathbf{A} = [1,0],\quad \mathbf{B} = [1,-1],\quad \mathbf{C} = [1, 1], \quad \mathbf{N} = [0,1], 
\end{align}
the (enhanced) global symmetry can be read off from the Kodaira classification  
\begin{align} 
\mathbf{A}_m &:\quad \mathbf{A}^{m+1},\cr
{\bf D}_{m\ge4} &:\quad \mathbf{A}^m\mathbf{B}\mathbf{C},\cr
{\bf E}_{m\ge6}&:\quad \mathbf{A}^{m-1}\mathbf{B}\mathbf{C}\mathbf{C},
\end{align}
up to $SL(2,\mathbb{Z})$ transformation. \\\hspace*{\fill}

\begin{figure}[t]
 \begin{center}
\includegraphics[width=13cm]{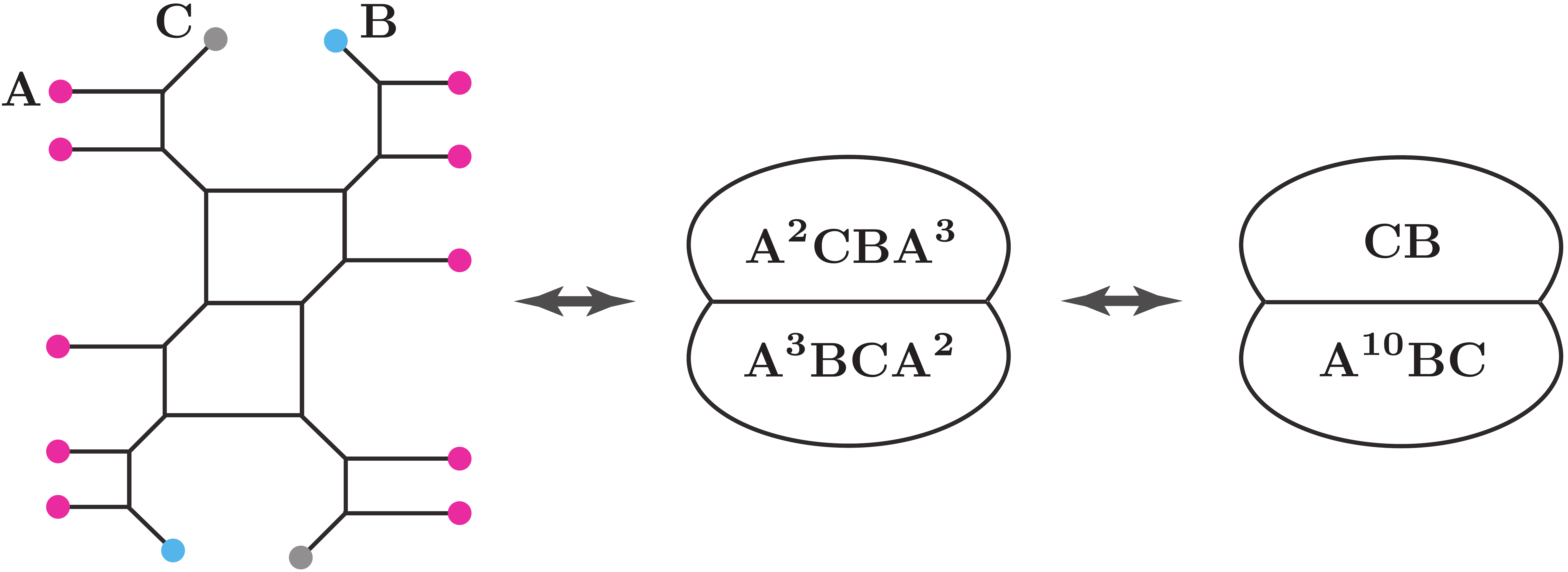}
 \end{center}
 \caption{A web diagram of $SU(3)$ theory with $N_f=10$ flavors in which 5-branes attached to 7-branes labeled by their charge vectors $\mathbf{A,B,C,N}$ is given on the left. The middle and right diagrams are the corresponding 7-brane configuration leading to the $\mathbf{D}_{10}=\mathbf{A}^{10}\mathbf{BC}$ corresponding to $SO(20)$ global symmetry.}
 \label{fig:nf10}
\end{figure}

%%%%%%%
\noindent\underline{\bf $\boldsymbol{SU(n)_0,N_f=2n+4}$   theory  }\\
Let us first start from the 5d $SU(n)$ gauge theory with $N_f = 2n+4$ and the vanishing CS level. An example of the web configuration for the $SU(3)$ theory with $N_f=10$ flavors is given in Figure \ref{fig:nf10}. 
In order to read off the global symmetry, we pull in all the 7-branes either inside the uppermost 5-brane loop or the lowermost the 5-brane loop. We will describe the 7-brane configuration as  ``${\mathbf({\rm lower}~ |~ {\rm upper})}$''. 
The configuration of 7-branes for the $SU(n)$ gauge theories with $N_f = 2n+4$ is then, 
by listing 7-branes counterclockwise from the left $n$ D7-branes, 
given by
\begin{align}\label{su3nf10mono}
(
\mathbf{A}^n\mathbf{BCA}^{2}|\mathbf{A}^n\mathbf{BCA}^{2}).
\end{align}
Recall that the position of a flavor brane leads to the mass parameter of the hypermultiplet and by mass deformation its position can be relocated through the different Coulomb moduli. D7-branes can cross D5-branes without affecting them, and this means that the configuration $\mathbf{A}$ is not confined within a 5-brane loop. Together with 7-brane monodromies, e.g., 
\begin{align}\label{basicmono}
\mathbf{AB} = \mathbf{BN},\quad \mathbf{NC} = \mathbf{CA},\quad \Rightarrow \quad
 \mathbf{ABC}= \mathbf{BCA},
\end{align}
one easily finds that the 7-brane configuration \eqref{su3nf10mono} becomes
\begin{align}
\mathbf{(A}^{2n+4}\mathbf{BC|BC)}.
\end{align}
Hence, this gives an $SO(4n+8)$ global symmetry which is enhanced from $U(2n+4)$ symmetry of $2n+4$ flavors.
It should be emphasized that Tao diagram contains spiral rotations associated with KK modes whose period is the instanton factor. The enhanced global symmetry for $N_f=2n+4$ flavors is then  
\begin{align}
SO(4n+8)\times U(1)_I  \supset U(2n+4)\times U(1)_I,
\end{align} 
where $U(1)_I$ is symmetry of instanton particle coming from $S^1$ compactification.

\begin{figure}[t]
 \begin{center}
\includegraphics[width=13cm]{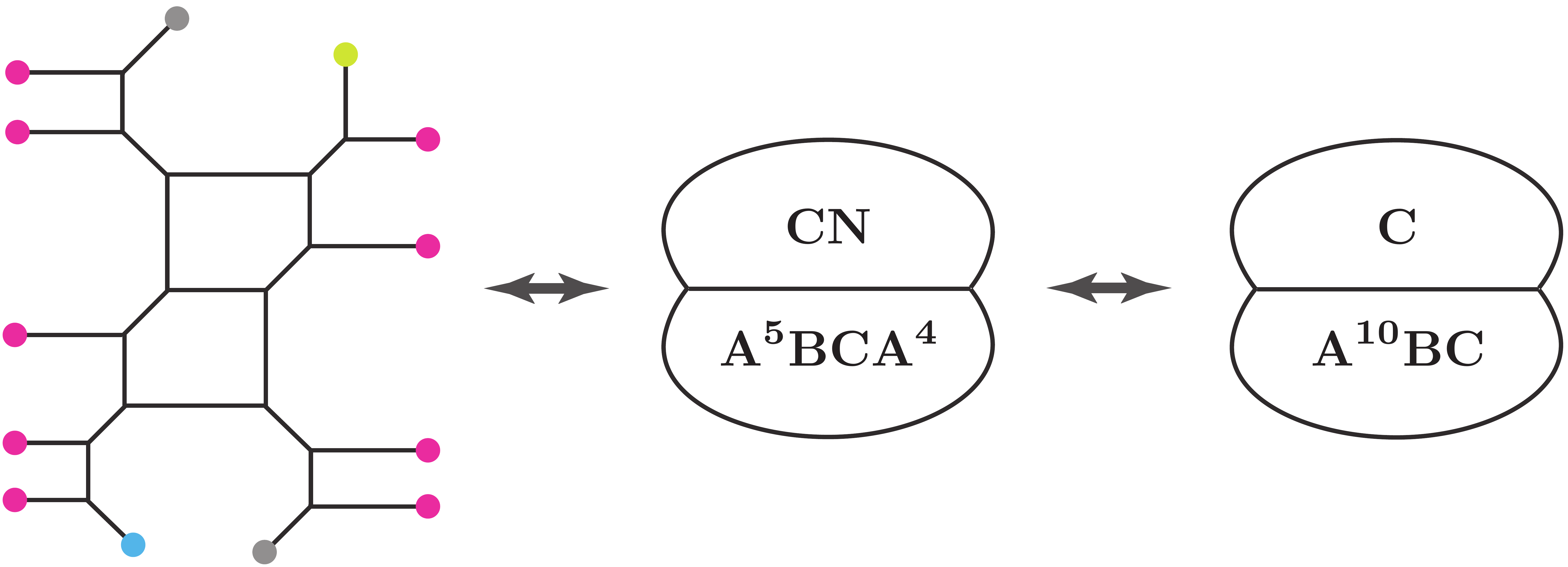}
 \end{center}
 \caption{A web diagram of $SU(3)$ theory with $N_f=9$ flavors in which 5-branes attached to 7-branes labeled by their charge vectors $\mathbf{A,B,C,D}$, and 7-brane configuration leading to the $\mathbf{D}_{10}=\mathbf{A}^{10}\mathbf{BC}$ corresponding to $SO(20)$ global symmetry.}
 \label{fig:nf9}
\end{figure}
\vspace{1cm}
\begin{table}
\begin{center}
\begin{tabular}{ c | l  }
  \hline
  $N_f$ &  $G_{|\kappa| }$ \\
    \hline
  $2n+4$ & ${SO}(4n+8)_{0}$ \\
  $2n+3$ & $SO(4n+8)_{\frac12}$ \\
  $2n+2$&$SU(2n+4)_{0},\qquad \big[SO(4n+4)\times SU(2)\big]_{1}$\\
  $2n+1$&$\big[SU(2n+2)\times SU(2)\big]_{\frac12},\qquad SO(4n+2)_{\frac32}$\\
  $2n$&$\big[SU(2n)\times SU(2)\times SU(2)\big]_{0},\qquad SU(2n+1)_{1},\qquad SO(4n)_{2}$\\
  \hline
  \end{tabular}
\end{center}
\caption{The non-abelian part of the enhanced global symmetries $G_{|\kappa|}$ of 5d $SU(n)_\kappa$ gauge theories with $N_f \geq 2n$ at UV fixed point.}
\label{tb:SU(n)}
\end{table}

%%%%%%%
\noindent\underline{\bf $\boldsymbol{SU(n)_\frac12, N_f=2n+3}$   theory  }\\
We then decouple one flavor and consider an $SU(n)$ gauge theory with $N_f = 2n+3$ and the CS level $\kappa = \pm \frac{1}{2}$. An example of the 5-brane web diagram for $SU(3)$ gauge theory with $N_f=9$ flavors is given in Figure \ref{fig:nf9} which can be understood as the mass decoupling limit where the position of one of flavor D5-branes is taken to infinity.
As can be read off from Figure \ref{fig:nf9}, it has the CS level $\kappa=-\frac12$. The web diagram is now no longer a Tao diagram and it is a usual web diagram of $(p, q)$ 5-branes. Hence, we expect that its UV completion is a 5d SCFT and it has an enhancement of the global symmetry at the fixed point. 
The global symmetry of the $SU(n)$ gauge theory with $N_f = 2n+3$ and $\kappa = -\frac{1}{2}$ is again dictated in the 7-brane monodromy 
\begin{align}
\mathbf{(A}^{n}\mathbf{BCA}^{2}|\mathbf{A}^{n-1}\mathbf{NCA}^{2}) = \mathbf{(A}^{n+2}\mathbf{BCA}^{n+1}|\mathbf{NC)} 
=  \mathbf{(A}^{2n+3}\mathbf{BC|CA)}=  \mathbf{(A}^{2n+4}\mathbf{BC|C)},
\end{align}
where we used \eqref{basicmono}, $\mathbf{NC=CA}$, which provides an additional $\mathbf{A}$ 7-brane thus yielding enhanced $SO(2n+4)$ symmetry
\begin{align}
SO(2n+4) \supset U(2n+3)\times U(1).
\end{align}
The $\kappa = \frac{1}{2}$ case 
also gives rise to the same enhanced global symmetry. As the CS level for 5d theories is subjected to \eqref{5dfixedptcond}, the allowed CS levels for $N_f=2n+3$ flavors are only $\kappa=\pm \frac12$. For less flavors, more CS levels are allowed and the corresponding global symmetry may have different enhanced global symmetry depending on the CS level.\\\hspace*{\fill}

\begin{table}[t]
\begin{center}
\begin{tabular}{ c | l  }
  \hline
  $N_f$ & $G_{|\kappa|}$ ($\kappa$ is the Chern-Simons level) \\
    \hline
  $10$ & ${SO}(20)_{0}$  \\
  $9$ & $SO(20)_{\frac12}$ \\
  $8$&$SU(10)_{0},\qquad \big[SO(16)\times SU(2)\big]_{1}$\\
  $7$&$\big[SU(8)\times SU(2)\big]_{\frac12},\qquad SO(14)_{\frac32}$\\
  $6$&$\big[SU(6)\times SU(2)\times SU(2)\big]_{0},\qquad SU(7)_{1},\qquad SO(12)_{2}$\\
  $5$&$\big[SU(5)\times SU(2)\big]_{\frac12},\qquad SU(6)_{\frac32},\qquad SO(10)_{\frac52}$\\
  $4$&$SU(4)_{0},\qquad \big[SU(4)\times SU(2)\big]_{1},\qquad SU(5)_{2},\qquad SO(8)_{3}$\\
  $3$&$ SU(3)_{\frac12},\qquad \big[SU(3)\times SU(2)\big]_{\frac32},\qquad SU(4)_{\frac52},\qquad SO(6)_{\frac72}$\\
  $2$&$SU(2)_{0},\qquad SU(2)_{1},\qquad \big[SU(2)\times SU(2)\big]_{2},\qquad SU(3)_3,\qquad SO(4)_4$\\
  $1$&$SU(2)_{\frac52},  \qquad  SU(2)_{\frac72}$\\
  $0$&$SU(2)_3$\\
  \hline
  \end{tabular}
\end{center}
\caption{The non-abelian part of the global symmetries of all the 5d $SU(3)$ gauge theories which have the UV completion. The theories that do not have non-abelian global symmetry are omitted.}
\label{tb:SU(3)}
\end{table}
%

%%%%%%%
\noindent\underline{\bf $\boldsymbol{SU(n)_\kappa,~2n\le N_f \le 2n+2}$  theory}\\ 
Using 7-brane monodromies, it is straightforward to generalize to $SU(n)$ gauge theories with $2n\le N_f\le 2n+2$ and the CS level which satisfies \eqref{5dfixedptcond} by decoupling the flavors one by one. The resulting theories are supposed to have the 5d UV fixed point. It is convenient to label the corresponding global symmetries by the CS level $|\kappa|$, like $G_{|\kappa|}$, and 
the enhanced global symmetries $G_{|\kappa|}$ of the theories are summarized in Table \ref{tb:SU(n)}.

We also included the cases with $N_f = 2n+3, 2n+4$ for completeness. In the Table \ref{tb:SU(n)}, we only wrote the non-abelian part of the enhanced global symmetries. The abelian part can be easily reproduced since the total number of the rank of the global symmetry should be $N_f  +1$, which corresponds to the number of the parameters of the theory. The theories exhaust all the possibilities with $N_f \geq 2n$ which satisfy \eqref{5dfixedptcond}, and all the theories show the enhancement of the global symmetries. \\\hspace*{\fill}

It is straightforward to check the global symmetry of $SU(n)_\kappa$ theory for lower flavors, and there are various global symmetries depending on the CS level $\kappa$ at a given number flavor. It is worth of noting that for $N_f=0,1$ cases, however, not all CS level lead to non-abelian global symmetry; it appears only for the following CS levels (denoted in the subscript):
\begin{align}
N_f= 0: SU(2)_{n},\qquad N_f= 1: SU(2)_{n-\frac12 }, \ SU(2)_{n+ \frac12 }.
\end{align}

%%%%%%%
\noindent\underline{\bf $\boldsymbol{SU(3)_\kappa, N_f}$   theory with $\boldsymbol{N_f \leq 10}$ flavors}\\
As a concrete example of the enhanced global symmetries for a given $SU(n)$ gauge theory, we pick $n=3$ and list all the global symmetries of 5d $SU(3)$ gauge theory of all the allowed flavors and the CS level $\kappa$ which has the UV completion in Table \ref{tb:SU(3)}.

We wrote down only the non-abelian part in Table \ref{tb:SU(3)}. There is again an abelian part such that the total rank of the global symmetry is $N_f+1$.   Bergman et. al. have found a considerable part of the list of the enhanced global symmetry~\cite{Bergman:2013aca, Bergman:2014kza}.  Here we enclose all for the completeness. Our results for $N_f=10,9$ and  a few higher Chern-Simons level    cases seem new.

\section*{Acknowledgements}
We thank  Amihay Hanany, Seok Kim, Ashoke Sen and David Tong for useful discussions.
The work of H.H. is supported by the grant FPA2012-32828 from the MINECO, the REA grant agreement PCIG10-GA-2011-304023 from the People Programme of FP7 (Marie Curie Action), the ERC Advanced Grant SPLE under contract ERC-2012-ADG-20120216-320421 and the grant SEV-2012-0249 of the ``Centro de Excelencia Severo Ochoa" Programme.
The work of K.L. is supported in part by the National Research Foundation of Korea (NRF) Grants No. 2006-0093850. The work of M.T. is supported by the RIKEN iTHES project.

%%%%%%%%%%%%%%%%%%%%%%%%%%%%%%%%%%%%%%%%%%%%%%%%%%
\providecommand{\href}[2]{#2}\begingroup\raggedright\endgroup

\end{document}